\documentclass[conference]{IEEEtran}
\IEEEoverridecommandlockouts
\usepackage{cite}
\usepackage{amsmath,amssymb,amsfonts}
\usepackage{algorithmic}
\usepackage{graphicx}
\usepackage{textcomp}
\usepackage{xcolor}
\usepackage{booktabs}
\usepackage{array} 
\usepackage{balance}
\usepackage{url}

\def\BibTeX{{\rm B\kern-.05em{\sc i\kern-.025em b}\kern-.08em
    T\kern-.1667em\lower.7ex\hbox{E}\kern-.125emX}}
\begin{document}

%\title{Language-Based Species Retrieval \\in Hypercube Embeddings\\
%\title{Querying the Wild: Language-Based Retrieval in Hypercube Embeddings
\title{Compact Hypercube Embeddings for Fast Text-based Wildlife Observation Retrieval

}

\author{\IEEEauthorblockN{Ilyass Moummad}
\IEEEauthorblockA{\textit{Inria, LIRMM, UM} \\
France \\
ilyass.moummad@inria.fr}
\and
\IEEEauthorblockN{Marius Miron}
\IEEEauthorblockA{\textit{Earth Species Project} \\
Spain \\
marius@earthspecies.org}
\and
\IEEEauthorblockN{David Robinson}
\IEEEauthorblockA{\textit{Earth Species Project} \\
Australia \\
david@earthspecies.org}
\and
\IEEEauthorblockN{Kawtar Zaher}
\IEEEauthorblockA{\textit{INA} \\
France \\
kawtar.zaher@ina.fr}
\and
\IEEEauthorblockN{Hervé Goëau}
\IEEEauthorblockA{\textit{AMAP, CIRAD} \\
France \\
herve.goeau@cirad.fr}
\and
\IEEEauthorblockN{Olivier Pietquin}
\IEEEauthorblockA{\textit{Earth Species Project} \\
Belgium \\
olivier@earthspecies.org}
\and
\IEEEauthorblockN{Pierre Bonnet}
\IEEEauthorblockA{\textit{AMAP, CIRAD} \\
France \\
pierre.bonnet@cirad.fr} 
\and
\IEEEauthorblockN{Emmanuel Chemla}
\IEEEauthorblockA{\textit{Earth Species Project} \\
France \\
emmanuel@earthspecies.org}
\and
\IEEEauthorblockN{Matthieu Geist}
\IEEEauthorblockA{\textit{Earth Species Project} \\
France \\
matthieu@earthspecies.org}
\and
\IEEEauthorblockN{Alexis Joly}
\IEEEauthorblockA{\textit{Inria, LIRMM, UM} \\
France \\
alexis.joly@inria.fr} 
}

% \author{
% \IEEEauthorblockN{FirstnameA LastnameA}
% \IEEEauthorblockA{\textit{Affiliation\_Author\_One} \\
% Country\_Author\_One \\
% email\_author\_one@domain.com}
% \and
% \IEEEauthorblockN{FirstnameB LastnameB}
% \IEEEauthorblockA{\textit{Affiliation\_Author\_Two} \\
% Country\_Author\_Two \\
% email\_author\_two@domain.com}
% \and
% \IEEEauthorblockN{FirstnameC LastnameC}
% \IEEEauthorblockA{\textit{Affiliation\_Author\_Three} \\
% Country\_Author\_Three \\
% email\_author\_three@domain.com}
% \and
% \IEEEauthorblockN{FirstnameD LastnameD}
% \IEEEauthorblockA{\textit{Affiliation\_Author\_Four} \\
% Country\_Author\_Four \\
% email\_author\_four@domain.com}
% \and
% \IEEEauthorblockN{FirstnameE LastnameE}
% \IEEEauthorblockA{\textit{Affiliation\_Author\_Five} \\
% Country\_Author\_Five \\
% email\_author\_five@domain.com}
% \and
% \IEEEauthorblockN{FirstnameF LastnameF}
% \IEEEauthorblockA{\textit{Affiliation\_Author\_Six} \\
% Country\_Author\_Six \\
% email\_author\_six@domain.com}
% }

\maketitle

\begin{abstract}
Large-scale biodiversity monitoring platforms increasingly rely on multimodal wildlife observations. While recent foundation models enable rich semantic representations across vision, audio, and language, retrieving relevant observations from massive archives remains challenging due to the computational cost of high-dimensional similarity search. In this work, we introduce \emph{compact hypercube embeddings for fast text-based wildlife observation retrieval}, a framework that enables efficient text-based search over large-scale wildlife image and audio databases using compact binary representations. Building on the \textit{cross-view code alignment} hashing framework, we extend lightweight hashing beyond single-modality setup to align natural language descriptions with visual or acoustic observations in a shared Hamming space. Our approach leverages pretrained wildlife foundation models, including BioCLIP, and BioLingual, and adapts them efficiently for hashing using parameter-efficient fine-tuning. We evaluate our method on large-scale benchmarks, including iNaturalist2024 for text-to-image retrieval and iNatSounds2024 for text-to-audio retrieval, as well as multiple soundscape datasets to assess robustness under domain shift. Results show that retrieval using the discrete hypercube embeddings achieves competitive—and in several cases superior—performance compared to the continuous embeddings, while drastically reducing memory and search cost. 
Moreover, we observe that the hashing objective consistently improves the underlying encoder representations, leading to stronger retrieval and zero-shot generalization. These results demonstrate that binary, language-based retrieval enables scalable and efficient search over large wildlife archives for biodiversity monitoring systems.\footnote{Code is available at: \url{https://github.com/ilyassmoummad/wildlifehashing}}
\end{abstract}


\section{Introduction}

%Motivation: retrieval at scale for biodiversity
Global biodiversity monitoring increasingly relies on massive collections of wildlife observations, including images and audio recordings, from camera traps~\cite{cameratraps}, passive acoustic monitoring~\cite{pam}, citizen science observation databases~\cite{fraisl2022citizen}. Platforms such as iNaturalist~\cite{inaturalist}, eBird~\cite{ebird}, and Pl@ntNet~\cite{plantnet} now host tens to hundreds of millions of observations spanning thousands of species, with new data continuously added. Beyond their sheer scale, these collections are inherently multimodal, combining visual, acoustic, and textual information. As a result, retrieving relevant observations from large and heterogeneous biodiversity archives has become a central challenge.

%From classification to retrieval
Most existing systems address biodiversity data through \emph{species classification}, predicting a single taxonomic label for an image or an audio recording~\cite{plantnet, birdnet}. While classification-based pipelines have achieved impressive accuracy, they are insufficient for many real-world biodiversity workflows. Ecologists may wish to retrieve all observations corresponding to a species using its scientific or common name, conservation practitioners may explore visually or acoustically similar observations, and citizen scientists may issue natural language queries describing organisms or behaviors rather than exact taxonomic labels~\cite{inquire}. These use cases call for \emph{retrieval-oriented} systems that support flexible, text-based querying over large multimodal wildlife data.

%Existing systems and their limitations
Several deployed tools, such as BirdNET~\cite{birdnet} for birds and Pl@ntNet~\cite{plantnet} for plants, demonstrate that accurate species identification can be achieved on resource-constrained devices. However, these systems are designed for closed-set classification rather than general-purpose retrieval. On the retrieval side, recent benchmarks have begun to emphasize retrieval as a more realistic paradigm for biodiversity monitoring. BIRB~\cite{birb} introduces an audio-to-audio retrieval task in bioacoustics, while INQUIRE~\cite{inquire} targets text-to-image retrieval for nature imagery. Although these efforts highlight the importance of retrieval, they rely on high-dimensional continuous embeddings and do not explicitly address the representation and storage constraints encountered in large-scale biodiversity platforms~\cite{tabmon}.
\looseness=-1

%Why efficiency and scalability matter
Scalability is a fundamental concern in biodiversity applications~\cite{scalingbiodiv}. Wildlife datasets are not only large, but are often accessed on mobile devices~\cite{plantnet} or processed by embedded sensors deployed in the field~\cite{tabmon}. Modern vision–language and audio–language foundation models produce highly expressive continuous embeddings, but storing and comparing such representations at scale remains challenging. As biodiversity archives continue to grow, representation choices become increasingly important for enabling practical retrieval systems.

%Hashing for efficient retrieval
Hashing provides a principled alternative by mapping continuous representations into compact binary codes that preserve semantic similarity~\cite{hashsurvey}. Binary representations enable retrieval using Hamming distance and allow similarity search to be implemented using simple bitwise operations. Hashing has been extensively studied in computer vision and cross-modal retrieval, yet it has received little attention in the context of biodiversity monitoring, despite its natural fit for large-scale and multimodal wildlife data.

%Our approach
In this work, we introduce a unified framework for \emph{text-based retrieval of wildlife observations in hypercube embeddings}. Building on the Cross-View Code Alignment (CroVCA) hashing framework~\cite{crovca}, we extend lightweight hashing beyond single-modality settings to align natural language descriptions with visual or acoustic observations in a shared Hamming space. Our approach leverages pretrained wildlife foundation models, including BioCLIP and BioLingual, and adapts them to binary retrieval using parameter-efficient fine-tuning. This design enables language-driven retrieval over large collections of wildlife images and sounds while preserving the semantic structure learned by pretrained multimodal encoders.
\looseness=-1

%Contrib
This work makes the following contributions:
\begin{itemize}
    \item We introduce the first framework for \textbf{text-based wildlife retrieval using binary hypercube embeddings}, enabling language-driven search over large-scale image and audio biodiversity datasets.
    \item We extend \textbf{cross-view code alignment} to a cross-modal setting, aligning textual descriptions with visual and acoustic observations in a shared Hamming space using lightweight hashing heads.
    \item We empirically show that training with a cross-modal hashing objective can \textbf{improve representation quality}, leading to stronger retrieval performance and improved zero-shot generalization under domain shift.
\end{itemize}

\section{Related Work}

Early work in computational biodiversity has primarily focused on species classification, where models predict a single taxonomic label for an image or audio recording~\cite{plantnet, birdnet}. While effective for identification, classification-centric approaches are limited in their ability to support exploratory analysis, open-ended search, and interaction with large biodiversity archives. As a result, recent work has increasingly emphasized \emph{retrieval} as a complementary paradigm for biodiversity monitoring~\cite{birb, inquire}.

In bioacoustics, the BIRB benchmark~\cite{birb} introduced a retrieval-oriented evaluation protocol by measuring a system’s ability to retrieve acoustically similar bird vocalizations rather than performing closed-set classification. This formulation better reflects ecological workflows such as validating detections and analyzing call variation, but focuses exclusively on audio-to-audio retrieval and relies on continuous embeddings. Beyond audio, INQUIRE~\cite{inquire} proposes a benchmark for text-to-image retrieval in the natural world, highlighting the potential of language-based interfaces for biodiversity exploration. However, retrieval is again performed using high-dimensional continuous representations, and scalability or representation constraints are not explicitly addressed. Existing benchmarks therefore either focus on a single modality or do not consider the representation challenges inherent to large-scale biodiversity platforms~\cite{tabmon}. In particular, text-to-audio retrieval for wildlife monitoring remains largely unexplored.

Recent advances in foundation models have significantly improved representation learning for biodiversity data~\cite{biolingual, bioclip, bioclip2, taxabind}. Vision–language models such as BioCLIP~\cite{bioclip} and BioCLIP-2~\cite{bioclip2} learn semantically rich embeddings from large-scale image–text supervision, while audio–language models such as BioLingual~\cite{biolingual} align bioacoustic recordings with textual descriptions, enabling zero-shot classification and retrieval. Although these models support cross-modal retrieval using cosine similarity, their reliance on high-dimensional continuous embeddings poses challenges for large and continuously growing biodiversity archives.

Hashing methods address approximate nearest neighbor search by mapping high-dimensional representations to compact binary codes while preserving semantic similarity~\cite{hashsurvey}. Prior work spans unsupervised, supervised, and cross-modal settings, but cross-modal hashing methods often rely on complex training objectives, modality-specific losses, or multi-stage optimization, making them difficult to adapt to large pretrained models~\cite{crossmodalhashsurvey}. More recently, Cross-View Code Alignment (CroVCA)~\cite{crovca} introduced a lightweight hashing framework in which augmented versions of the same image are encouraged to share identical hash codes, while representational collapse is avoided using a coding-rate regularization objective. In this work, we build on these principles and extend cross-view code alignment beyond single-modality settings to align natural language descriptions with visual or acoustic wildlife observations in a shared Hamming space.

\section{Method}
\label{sec:method}

\begin{figure*}[h]
  \centering
  \includegraphics[width=0.75\textwidth]{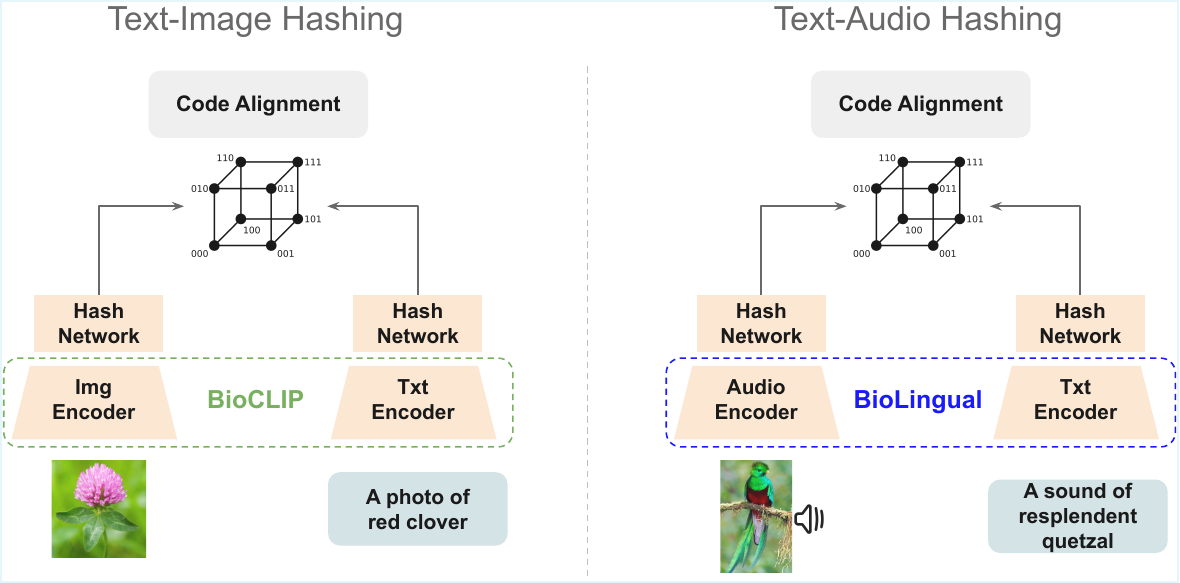}
  \caption{Overview of the proposed text--observation hashing framework for wildlife retrieval. Textual species descriptions and wildlife observations (images or audio) are encoded into compact binary codes and aligned using cross-view code alignment.}
  \label{fig:pipeline}
\end{figure*}

Building on the principles of \textit{cross-view code alignment}~\cite{crovca}, we propose a \emph{text–observation hashing} framework to enable efficient language-based retrieval of wildlife observations. Our method learns compact binary codes that align textual species descriptions with image or audio observations, in a shared Hamming hypercube.

\subsection{Problem Formulation}

Let $\mathcal{X}^{\text{text}}$ denote the space of textual descriptions (e.g., scientific or common species names) and $\mathcal{X}^{\text{obs}}$ denote the space of wildlife observations, where each observation $x^{\text{obs}}$ is either an image or an audio recording.  

Given a paired sample $(x^{\text{text}}, x^{\text{obs}})$ representing the same species, our goal is to learn modality-specific hashing functions
\looseness=-1
\begin{align}
\phi^{\text{text}} &: \mathcal{X}^{\text{text}} \rightarrow \{0,1\}^b, \\
\phi^{\text{obs}} &: \mathcal{X}^{\text{obs}} \rightarrow \{0,1\}^b,
\end{align}
that map text and observation inputs into a shared $b$-bit Hamming space. These mappings are designed such that the binary code of an observation matches the code of its corresponding textual description.  

At inference, language-based retrieval reduces to finding observations whose binary codes are closest in Hamming distance to the code generated from a text query:
\begin{equation}
x^{\text{obs}}_{\text{retrieved}} = \arg \min_{x^{\text{obs}}} d_H \big(\phi^{\text{text}}(x^{\text{text}}), \phi^{\text{obs}}(x^{\text{obs}})\big),
\end{equation}
where $d_H$ denotes Hamming distance. This formulation explicitly models separate encoders for text and observations while enforcing cross-modal alignment in the Hamming hypercube.

\subsection{Learning To Hash}

The model consists of modality-specific encoders followed by lightweight hashing networks, projecting features into the $b$-bit Hamming hypercube (Figure~\ref{fig:pipeline}).  

\paragraph{Text Encoder and Hashing Head.}  
A textual input $x^{\text{text}}$ is first embedded using a language encoder $f^{\text{text}}$:
\begin{equation}
h^{\text{text}} = f^{\text{text}}(x^{\text{text}}) \in \mathbb{R}^{d}.
\end{equation}
The resulting feature is then passed through a shallow hashing network $g^{\text{text}}$ to produce pre-binarized logits:
\begin{equation}
z^{\text{text}} = g^{\text{text}}(h^{\text{text}}) \in \mathbb{R}^{b}.
\end{equation}
Finally, the logits are squashed via a sigmoid function to produce probabilities and binarized to obtain $b$-bit codes:
\begin{equation}
p^{\text{text}} = \sigma(z^{\text{text}}), \quad
y^{\text{text}} = \mathbb{I}[p^{\text{text}} \ge 0.5].
\end{equation}

\paragraph{Observation Encoder and Hashing Head.}  
Similarly, for a wildlife observation $x^{\text{obs}}$, modality-specific encoders $f^{\text{obs}}$ are used: BioCLIP's image encoder for images or BioLingual's audio encoder for sound recordings. Features are projected via the corresponding hashing network $g^{\text{obs}}$:
\begin{equation}
h^{\text{obs}} = f^{\text{obs}}(x^{\text{obs}}), \quad
z^{\text{obs}} = g^{\text{obs}}(h^{\text{obs}}),
\end{equation}
and binarized similarly:
\begin{equation}
p^{\text{obs}} = \sigma(z^{\text{obs}}), \quad
y^{\text{obs}} = \mathbb{I}[p^{\text{obs}} \ge 0.5].
\end{equation}

\subsection{Text–Observation Code Alignment}

We treat textual and observation modalities as two semantic views of the same wildlife species. To enforce semantic consistency, we align their binary codes using a symmetric \emph{binary cross-entropy} (BCE) loss:
\begin{equation}
\mathcal{L}_{\text{align}} =
\frac{1}{2} \Big(
\mathrm{BCE}(y^{\text{text}}, p^{\text{obs}}) + 
\mathrm{BCE}(y^{\text{obs}}, p^{\text{text}})
\Big),
\label{eq:align}
\end{equation}
where the gradient is stopped on the binarized codes $y^{\text{text}}$ and $y^{\text{obs}}$ to encourages the predicted probability distributions of one modality to match the binarized representation of the other, promoting cross-view code alignment.

\subsection{Anti-Collapse Regularization}

Eq.~\eqref{eq:align} is not enough to learn meaningful codes, a common failure mode in representation learning is \emph{representation collapse}, where all inputs map to the same output. To mitigate this, we adopt a \emph{Maximum Coding Rate} (MCR) regularizer from CrovCA \cite{crovca}, which encourages feature diversity.

Given a batch of logits $\{z_i\}_{i=1}^B$, we normalize each vector:
\begin{equation}
v_i = \frac{z_i}{\|z_i\|_2}, \quad
C = \frac{1}{B} \sum_{i=1}^{B} v_i v_i^\top,
\end{equation}
and define the diversity regularizer:
\begin{equation}
\mathcal{L}_{\text{reg}} = - \frac{1}{2} \log \det \Big(I + \frac{b}{B} C \Big).
\end{equation}
This term encourages the logits to occupy all available dimensions evenly, preventing trivial solutions and ensuring balanced bit usage.

We apply the regularizer independently to text and observation logits, yielding
\begin{equation}
\mathcal{L}_{\text{div}} = \frac{1}{2} \big( \mathcal{L}_{\text{reg}}(z^{\text{text}}) + \mathcal{L}_{\text{reg}}(z^{\text{obs}}) \big).
\end{equation}

\subsection{Overall Objective}

The total training loss combines cross-view alignment and diversity regularization:
\begin{equation}
\mathcal{L} = \mathcal{L}_{\text{align}} + \lambda \, \mathcal{L}_{\text{div}},
\end{equation}
where $\lambda$ balances semantic alignment with code diversity. The resulting model produces compact binary codes that are both semantically meaningful and maximally informative.

\subsection{Language-Based Wildlife Retrieval}

At inference, we first compute binary codes independently for text queries and all wildlife observations using the trained modality-specific hashing functions $\phi^{\text{text}}$ and $\phi^{\text{obs}}$.  

Given a text query $x^{\text{text}}$, retrieval consists of selecting observations whose binary codes are closest in Hamming space:
\begin{equation}
\mathcal{R}(x^{\text{text}}) = \big\{ x^{\text{obs}} \,\big|\, d_H \big( \phi^{\text{text}}(x^{\text{text}}), \phi^{\text{obs}}(x^{\text{obs}}) \big) \text{ is minimal} \big\},
\end{equation}
where $d_H(\cdot,\cdot)$ denotes the Hamming distance. In other words, the retrieved set $\mathcal{R}(x^{\text{text}})$ contains the wildlife observations whose binary codes are most similar to the query code.  

This approach enables highly efficient, memory-light retrieval, as Hamming distances can be computed extremely fast using bitwise operations, while still preserving semantic alignment between textual species descriptions and the corresponding wildlife observations.  

\section{Experiments}

We evaluate our text–observation hashing framework on large-scale wildlife datasets, including both image and audio observations. In all experiments, the hashing heads are trained on paired text–observation data to align binary codes across modalities, while the backbone encoders are finetuned using the parameter-efficient LoRA approach \cite{lora}. We conduct experiments with both 128-bit and 256-bit hash codes.

\subsection{Datasets and Experimental Setup}

For image-text hashing, we train on iNaturalist 2021 and test on iNaturalist 2024. We evaluate retrieval separately within each biological supercategory (amphibians, animalia, arachnids, insects, birds, mollusks, ray-finned fishes, and reptiles) to enable a more fine-grained assessment of retrieval performance within a taxonomically coherent group.
% , following standard supercategory splits summarized in Table~\ref{tab:inat21_codes}. 
Text–image alignment is learned using BioCLIP and BioCLIP-2 encoders, with textual descriptions derived from species annotations.

% \begin{table}[h]
% \centering
% \caption{iNaturalist image supercategory codes.}
% \begin{tabular}{ll}
% \toprule
% Code & Supercategory \\
% \midrule
% Amphi & Amphibians \\
% Anima & Animalia \\
% Arach & Arachnids \\
% Insec & Insects \\
% Mamml & Mammals \\
% Molls & Mollusks \\
% RayFi & Ray-finned Fishes \\
% Reptl & Reptiles \\
% \bottomrule
% \end{tabular}
% \label{tab:inat21_codes}
% \end{table}

For audio, we train on the iNatSounds 2024 training split~\cite{inatsounds}. In-distribution performance is evaluated on the test split, while out-of-distribution (OOD) generalization is assessed using multiple soundscape datasets. These datasets span a wide range of geographic regions and ecological contexts, including tropical forests (PER~\cite{per}, NES~\cite{nes}), island and endangered-species habitats (UHH~\cite{uhh}), high-elevation montane environments (HSN~\cite{hsn}, SNE~\cite{sne}), and temperate woodland soundscapes (SSW~\cite{ssw}). Dataset selection follow established soundscape benchmarking protocols~\cite{birdset}. 

Audio–text alignment is learned using BioLingual test and audio encoders. We evaluate retrieval performance using mean average precision (mAP@1000). For hashed representations, retrieval uses Hamming distance, while continuous embeddings use cosine similarity.

\subsection{Image Retrieval Results}

Table~\ref{tab:inat24_results} presents text-to-image retrieval results. 128-bit hashing leads to a substantial performance drop, while 256-bit codes achieve retrieval performance comparable to—or better than—the original model with continuous embeddings, highlighting the efficiency of cross-modal hashing. In most scenarios, LoRA fine-tuning improves cosine-based retrieval, indicating that the cross-modal hashing objective enhances the encoder’s internal representations. %Table~\ref{tab:inat24_results} confirms similar trends on iNaturalist 2024.

\begin{table*}[h]
\centering
\caption{Text-to-image retrieval mAP@1000 on the iNat2024 image validation set, split by supercategory. We compare continuous embeddings from the pretrained encoders, hash codes (128/256-bit), and continuous embeddings from encoders after fine-tuning.}
\resizebox{\textwidth}{!}{%
\begin{tabular}{l|ccccccccccc|c}
\toprule
\textbf{Feature} & \textbf{Amphi} & \textbf{Anima} & \textbf{Arach} & \textbf{Birds} & \textbf{Fungi} & \textbf{Insec} & \textbf{Mamml} & \textbf{Molls} & \textbf{Plant} & \textbf{RayFi} & \textbf{Reptl} & AVG \\
\midrule
\multicolumn{13}{c}{\textbf{BioCLIP}} \\
\midrule
Continuous (Pretrained) & 49.51 & 62.76 & 59.84 & 70.77 & 72.72 & 70.52 & 60.58 & 54.28 & 47.94 & 56.05 & 39.45 & 58.58 \\
\midrule
\multicolumn{13}{l}{128-bit Hashing} \\
Binary (Hashing) & 8.42 & 23.34 & 17.54 & 37.49 & 38.97 & 43.59 & 24.95 & 22.95 & 45.73 & 24.13 & 15.17 & 27.48 \\
Continuous (LoRA) & 54.10 & 67.72 & 64.72 & 73.89 & 79.63 & 75.99 & 64.79 & 59.94 & 53.42 & 62.29 & 44.42 & 63.71 \\
\midrule
\multicolumn{13}{l}{256-bit Hashing} \\
Binary (Hashing) & 43.29 & 50.98 & 44.36 & 63.41 & 69.33 & 71.18 & 52.10 & 51.16 & 66.46 & 63.17 & 39.51 & 55.90 \\
Continuous (LoRA) & 56.20 & 68.01 & 66.27 & 75.64 & 79.49 & 78.06 & 66.26 & 61.24 & 56.18 & 64.45 & 48.00 & 65.43 \\
\midrule
\multicolumn{13}{c}{\textbf{BioCLIP-2}} \\
\midrule
Continuous (Pretrained) & 63.74 & 56.97 & 45.48 & 89.43 & 61.85 & 66.89 & 86.60 & 51.09 & 49.58 & 76.09 & 67.96 & 65.06 \\
\midrule
\multicolumn{13}{l}{128-bit Hashing} \\
Binary (Hashing) & 32.78 & 59.22 & 58.29 & 79.03 & 79.02 & 83.25 & 69.32 & 63.25 & 79.09 & 67.12 & 47.33 & 65.24 \\
Continous (LoRA) & 58.62 & 55.38 & 45.77 & 88.01 & 62.81 & 67.39 & 83.99 & 50.45 & 52.00 & 74.49 & 63.31 & 63.83 \\
\midrule
\multicolumn{13}{l}{256-bit Hashing} \\
Binary (Hashing) & 71.35 & 81.07 & 86.28 & 88.58 & 88.15 & 92.53 & 86.01 & 85.53 & 88.38 & 84.50 & 73.21 & 84.14 \\
Continuous (LoRA) & 65.02 & 57.47 & 46.50 & 87.75 & 64.91 & 69.71 & 83.57 & 53.64 & 55.17 & 75.55 & 65.16 & 65.85 \\
\bottomrule
\end{tabular}%
}
\label{tab:inat24_results}
\end{table*}

Beyond retrieval accuracy, binary embeddings offer major efficiency gains: a 256-bit hash is 96$\times$ smaller than a 768-D float32 BioCLIP-2 embedding, and Hamming distance replaces $\sim 10^3$ floating-point operations per comparison with a fewer bitwise operations, enabling faster retrieval.

% Beyond retrieval accuracy, binary embeddings offer major computational and memory benefits. A BioCLIP-2 embedding has 768 float32 dimensions (3,072 bytes), whereas a 256-bit hash requires only 32 bytes, yielding a 96× reduction in storage. Cosine similarity requires $~10^3$ floating-point operations per comparison, while Hamming distance can be computed with a few bitwise operations, resulting in orders-of-magnitude faster retrieval.

\subsection{Audio Retrieval Results}

Table~\ref{tab:inatsounds24_results} reports text-to-audio retrieval results on iNatSounds 2024. 256-bit hashing provides the best trade-off between compression and retrieval accuracy. LoRA fine-tuning consistently improves performance over the original BioLingual encoder under cosine similarity.

\begin{table}[h]
\centering
\caption{Text-to-audio retrieval mAP@1000 on the iNatSounds 2024 validation set, split by supercategory. Results are shown for original embeddings, hashing, and after fine-tuning.}
\resizebox{\columnwidth}{!}{%
\begin{tabular}{l|ccccc|c}
\toprule
\textbf{Feature} & \textbf{Amphi} & \textbf{Aves} & \textbf{Insec} & \textbf{Mamml} & \textbf{Rept} & AVG \\
\midrule
BioLingual & 39.06 & 48.54 & 37.84 & 51.59 & 93.11 & 45.02 \\
\midrule
\multicolumn{7}{l}{128-bit Hashing} \\
Binary (Hashing) & 44.13 & 52.98 & 39.45 & 42.63 & 88.35 & 44.59 \\
Continous (LoRA) & 46.51 & 55.65 & 44.21 & 54.11 & 96.29 & 49.46 \\
\midrule
\multicolumn{7}{l}{256-bit Hashing} \\
Binary (Hashing) & 48.49 & 55.87 & 49.90 & 46.92 & 90.51 & 48.61 \\
Continuous (LoRA) & 50.05 & 58.21 & 50.30 & 59.28 & 94.64 & 52.08 \\
\bottomrule
\end{tabular}%
}
\label{tab:inatsounds24_results}
\end{table}

We assess domain generalization using soundscape datasets that differ from the training distribution (Table~\ref{tab:soundscape_results_hashing}). Both 128-bit and 256-bit hashed representations outperform the original BioLingual model, indicating that the hypercube bottleneck encourages more discriminative and robust features.

\begin{table}[h]
\centering
\caption{Out-of-distribution performance on soundscape datasets (HSN, NES, SNE, UHH, PER, SSW).}
\resizebox{\columnwidth}{!}{%
\begin{tabular}{l|cccccc|c}
\toprule
\textbf{Feature} & \textbf{HSN} & \textbf{NES} & \textbf{SNE} & \textbf{UHH} & \textbf{PER} & \textbf{SSW} & AVG \\
\midrule
BioLingual & 27.85 & 24.32 & 21.49 & 32.00 & 9.73 & 33.64 & 24.83 \\
\midrule
\multicolumn{8}{l}{128-bit Hashing} \\
%\midrule
Binary (Hashing) & 35.20 & 28.28 & 34.00 & 29.77 & 6.89 & 55.30 & 31.57 \\
Continuous (LoRA) & 40.69 & 29.94 & 30.48 & 28.30 & 9.96 & 49.15 & 31.42 \\
\midrule
\multicolumn{8}{l}{256-bit Hashing} \\
%\midrule
Binary (Hashing) & 39.65 & 28.80 & 34.02 & 29.66 & 8.70 & 55.75 & 32.76 \\
Continuous (LoRA) & 38.11 & 28.26 & 28.91 & 33.00 & 11.10 & 43.84 & 30.53 \\
\bottomrule
\end{tabular}%
}
\label{tab:soundscape_results_hashing}
\end{table}

To evaluate transferability, we perform zero-shot classification on the soundscape datasets (Table~\ref{tab:soundscape_results_zs}). Encoders fine-tuned with 128-bit hashing achieve the highest OOD accuracy, substantially improving over the original model. This demonstrates that hashing can enhance representation quality beyond retrieval tasks.

\begin{table}[h]
\centering
\caption{Zero-shot classification performance on soundscape OOD datasets, comparing the original and LoRA fine-tuned model.}
\resizebox{\columnwidth}{!}{%
\begin{tabular}{l|cccccc|c}
\toprule
\textbf{Feature} & \textbf{HSN} & \textbf{NES} & \textbf{SNE} & \textbf{UHH} & \textbf{PER} & \textbf{SSW} & AVG \\
\midrule
BioLingual & 19.69 & 32.18 & 42.70 & 16.65 & 7.68 & 35.98 & 25.81 \\
\midrule
\multicolumn{8}{l}{128-bit Hashing} \\
%\midrule
Continuous (LoRA) & 23.17 & 38.94 & 50.79 & 18.10 & 10.71 & 42.59 & 30.71 \\
\midrule
\multicolumn{8}{l}{256-bit Hashing} \\
%\midrule
Continuous (LoRA) & 20.83 & 30.85 & 41.68 & 20.34 & 11.41 & 41.21 & 27.72 \\
\bottomrule
\end{tabular}%
}
\label{tab:soundscape_results_zs}
\end{table}

\subsection{Summary}

Across image and audio modalities, text–observation hashing enables efficient retrieval without sacrificing accuracy. In many cases, particularly with BioCLIP-2, 256-bit hashed retrieval matches or exceeds continuous embedding performance, reducing memory and search costs. Moreover, the hashing objective consistently improves encoder representations, yielding stronger retrieval and zero-shot generalization.

\section{Conclusion}

We presented a unified framework for \emph{text-based wildlife retrieval in hypercube embeddings}, enabling scalable and efficient search over large-scale multimodal wildlife datasets. By extending the cross-view code alignment hashing paradigm to align textual descriptions with images and audio recordings in a shared Hamming space, our method produces compact binary representations that reduce retrieval cost. Experimental results on several benchmarks demonstrate that 256-bit hashing achieves retrieval performance comparable to or exceeding continuous embeddings. Furthermore, we observed that training with the cross-modal hashing objective improves the underlying encoder representations, enhancing zero-shot generalization and robustness under domain shift. By bridging the gap between efficient retrieval and high-quality representation learning, our approach lays the groundwork for interactive exploration of massive wildlife archives and opens new avenues for citizen science, ecological research, and conservation applications.

\section*{Acknowledgment}

This work was supported by the Pl@ntAgroEco project, funded by the "Agence Nationale de la Recherche" (ANR) under the France 2030 program, within the “Agroécologie et Numérique” initiative (reference ANR-22-PEAE-0009). The authors gratefully acknowledge this support.

% \section*{Acknowledgment}

% The preferred spelling of the word ``acknowledgment'' in America is without 
% an ``e'' after the ``g''. Avoid the stilted expression ``one of us (R. B. 
% G.) thanks $\ldots$''. Instead, try ``R. B. G. thanks$\ldots$''. Put sponsor 
% acknowledgments in the unnumbered footnote on the first page.

%\pagebreak
\bibliographystyle{IEEEtran}
\balance
\bibliography{refs}

\end{document}